\documentclass[aps,prl,reprint,superscriptaddress,nofootinbib,longbibliography]{revtex4-2}
\usepackage{amsmath,amssymb,bm}
\usepackage{graphicx}
\usepackage{booktabs}
\usepackage{microtype}
\usepackage{xcolor}
\usepackage{epsfig}
\usepackage{verbatim}
\usepackage{dcolumn}
\usepackage[colorlinks=true,citecolor=blue!55!black,linkcolor=blue!55!black,urlcolor=blue!55!black]{hyperref}

\usepackage{braket}
\usepackage{float}
\usepackage{enumitem}
\usepackage{multirow}
\usepackage{orcidlink}

\newcommand{\Tr}{\operatorname{Tr}}
\newcommand{\Diss}{\mathcal{D}}
\newcommand{\Resp}{\mathcal{F}}
\newcommand{\rhoU}{\rho_{\mathrm U}}
\newcommand{\gpop}{\gamma_{\mathrm p}}
\newcommand{\gcoh}{\gamma_{\mathrm c}}
\newcommand{\atanh}{\operatorname{artanh}}

\begin{document}

\title{Relativistic Strong Quantum Mpemba Effect}

	\author{Zhilong Liu~\orcidlink{0009-0000-0353-5113}}
	\affiliation{Department of Physics, Key Laboratory of Low Dimensional Quantum Structures and Quantum Control of Ministry of Education, and Synergetic Innovation Center for Quantum Effects and Applications, Hunan Normal University, Changsha, Hunan 410081, P. R. China}
		\author{Zehua Tian}
	\email{tzh@hznu.edu.cn } 
	\affiliation{School of Physics, Hangzhou Normal University, Hangzhou, Zhejiang 311121, China}
	
	\author{Hui Jing}
	\email{jinghui73@foxmail.com}
	\affiliation{Department of Physics, Key Laboratory of Low Dimensional Quantum Structures and Quantum Control of Ministry of Education, and Synergetic Innovation Center for Quantum Effects and Applications, Hunan Normal University, Changsha, Hunan 410081, P. R. China}

	\author{Jieci Wang~\orcidlink{0000-0001-5072-3096}}
	\email{jcwang@hunnu.edu.cn}
	\affiliation{Department of Physics, Key Laboratory of Low Dimensional Quantum Structures and Quantum Control of Ministry of Education, and Synergetic Innovation Center for Quantum Effects and Applications, Hunan Normal University, Changsha, Hunan 410081, P. R. China}

\begin{abstract}
The Unruh effect links acceleration to thermalization, providing a relativistic setting for anomalous quantum relaxation. We demonstrate a  {\it relativistic}  strong quantum Mpemba effect in the Unruh thermalization of a two-level detector coupled to a massless scalar field.
For a fixed pair of  isospectral initial states, relaxation curves that never cross in inertial motion exhibit a Mpemba crossing above a threshold acceleration. At a higher, precisely tuned acceleration, the initially farther state has zero overlap with the slowest decay mode and relaxes with a faster asymptotic exponential rate. We identify how competition between transition and dephasing channels governs the relaxation spectrum, where sufficiently strong longitudinal coupling enables acceleration to switch the relaxation bottleneck between population and coherence modes. We derive the spectral-switching condition and construct isospectral preparation protocols that eliminate slow-mode contributions within their physical domains. Our results establish acceleration as a control parameter for anomalous quantum relaxation, connecting relativistic quantum dynamics with nonequilibrium thermodynamics.	
\end{abstract}

\maketitle

\emph{Introduction.---}
\begin{figure}[t]
	\includegraphics[width=\linewidth]{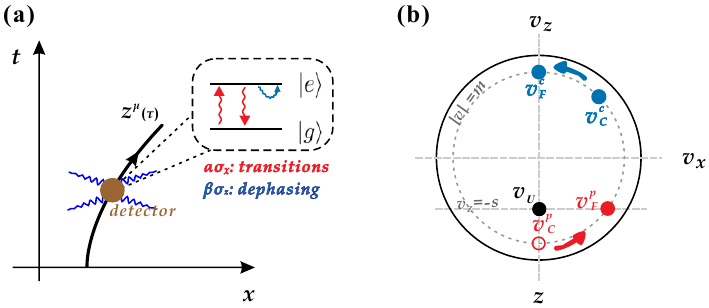}
	\caption{\textbf{Model diagram and	strong-Mpemba state preparation.}
		(a) A uniformly accelerated two-level detector follows the worldline $z^\mu(\tau)$ and interacts with the scalar field. (b) Isospectral state preparation on a fixed-purity Bloch orbit	$|\boldsymbol v|=m$. Both in the population-limited regime (red) and the coherence-limited regime (blue), the unitary rotation $\bm v_C\rightarrow \bm v_F$  eliminates the slowest mode weight. Open and filled circles denote the reference and optimized initial states, respectively.}
	\label{fig:sch}
\end{figure}
Although exposed to the same thermal environment, an initially hotter system can relax faster than a colder one, a phenomenon known as the Mpemba effect~\cite{EBMpemba:1969}. This anomalous phenomenon implies the counterintuitive fact that the distance to the thermal equilibrium state alone does not determine the relaxation time. For classical Markovian dynamics, Lu and Raz identified a spectral mechanism: an initially more distant state can have a smaller contribution from the slowest relaxation mode~\cite{LuRaz:2017}. In recent years, studies of this counterintuitive phenomenon have extended to both open and closed quantum systems~\cite{Bao:2025ocj,Beato:2025dbm,Strachan:2024mva,Bao:2022ffs,VanVu:2024zps,Ares:2022koq,Yamashika:2025vpd,Yu:2025lku,Yu:2025vth,Bertini:2022srv,Liu:2024kzv,Liu:2026rpz,Benjadi:2026lac,Xu:2026muy,Turkeshi:2024juo,Wang:2024jlr,Chatterjee:2023lgx,Chatterjee:2023lha,Wei:2026kub,Wei:2025uhv,Rylands:2023yzx}, including the experimental observation of quantum Mpemba relaxation in controlled quantum systems~\cite{AharonyShapira:2024nrt,Xu:2025wml,Joshi:2024sup,Zhang2025}. In Markovian open quantum systems, the strong quantum Mpemba effect involves eliminating the initial-state contribution from the slowest decaying modes, allowing faster modes to govern the asymptotic relaxation and thus yielding a faster exponential approach to equilibrium~\cite{Carollo2021}. This connection between initial-state preparation and relaxation modes raises the question of how relativistic motion can influence the emergence of the effect.

Unruh thermalization provides a setting  to investigate how relativistic motion controls relaxation. In the weak-coupling regime, a uniformly accelerated detector interacting with a quantum field in the Minkowski vacuum approaches a Gibbs state at the Unruh temperature~\cite{Unruh1976,Fulling:1972md,Davies:1974th}. Acceleration determines this stationary state and modifies the field correlations governing the detector’s relaxation. Wang et al. recently examined fidelity-based heating--cooling asymmetries and relaxation from equal-fidelity nonthermal states in Unruh thermalization, describing these behaviors as Mpemba-like~\cite{Wang:2025mpu}. These findings motivate a distinct question: can acceleration itself serve as a control parameter that induces a Mpemba crossing or even a strong one?

In this Letter, we show that acceleration can control the emergence of a relativistic strong quantum Mpemba effect in Unruh thermalization. For a fixed pair of suitably chosen isospectral states and sufficiently strong longitudinal coupling, increasing acceleration induces a relaxation crossing that is absent in inertial motion. At a second precisely matched acceleration, exact cancellation of the slow-mode contribution from the initially farther state produces the strong Mpemba effect. We derive the conditions for this sequence, identify the switch between population- and coherence-limited relaxation, and construct the corresponding isospectral preparation protocols. These constructions clarify how acceleration can suppress or enhance slow-mode overlap under a fixed-purity constraint, linking its role in relaxation to the underlying spectral structure. 

\emph{Generalized accelerated-detector dynamics.-}
Working in units $\hbar=c=k_B=1$, we consider the two-level detector illustrated in Fig.~\ref{fig:sch}(a), with Hamiltonian $H_{\mathrm D}=\omega_0\sigma_z/2$ and energy gap $\omega_0>0$. The detector undergoes uniform proper acceleration $a$ along the $x$ direction. Its worldline in
coordinates $z^\mu=(t,x,y,z)$ is~\cite{Kolekar:2013hra}
\begin{equation}
	z^\mu(\tau)=a^{-1}\bigl(\sinh(a\tau),\cosh(a\tau),0,0\bigr),
	\label{eq:worldline}
\end{equation}
where $\tau$ denotes the detector proper time. We then consider a generalized interaction between the detector and a massless scalar field~\cite{Liu:2025bzv,Liu:2025rks}
\begin{equation}
	H_{\mathrm I}(\tau)
	=g\bigl(\alpha\sigma_x+\beta\sigma_z\bigr)
	\otimes\hat{\phi}[z^\mu(\tau)].
	\label{eq:interaction}
\end{equation}
Here $g$ is the overall weak detector-field coupling. The real dimensionless coefficients $\alpha$ and $\beta$ weight the transverse transition and longitudinal dephasing couplings, respectively.
\begin{figure}[t]
	\includegraphics[width=0.8\linewidth]{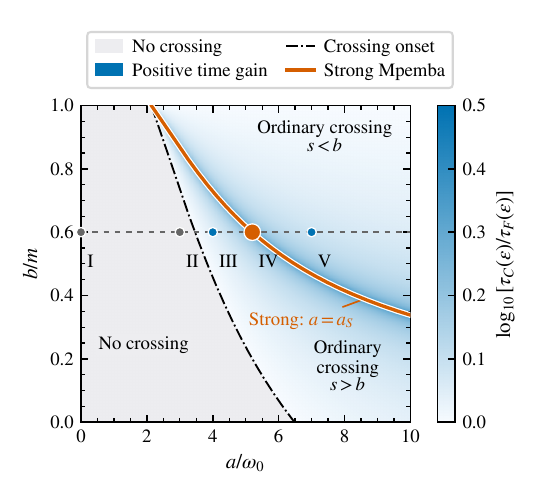}
	\caption{\textbf{Emergence of strong Mpemba effect.}
		Blue shading shows \(\log_{10}[\tau_C(\varepsilon)/\tau_F(\varepsilon)]\), where \(\tau_i\) is the first time at which \(D_T^i\le\varepsilon\). Gray denotes no crossing. The black dash-dotted curve marks crossing onset; the orange curve \(b=s\) marks exact slow-mode cancellation. The horizontal cut is at \(b/m=0.6\). m=0.9, \(\kappa=2\), and \(\varepsilon=10^{-4}\).}
	\label{fig:ESME}
\end{figure}

For uniform acceleration in the Minkowski vacuum, the positive-frequency Wightman function,
$G_a^+(s)=\langle 0_{\mathrm M}|\hat{\phi}[z(\tau)]	\hat{\phi}[z(\tau-s)]	|0_{\mathrm M}\rangle $, depends only on the proper-time difference $s$. The corresponding response functions are~\cite{Birrell:1982ix}
\begin{align}
	\Resp_a(\omega)
	&=\int_{-\infty}^{\infty}\!ds\,
	e^{-i\omega s}G_a^+(s)
	=\frac{\omega}{2\pi}
	\frac{1}{e^{2\pi\omega/a}-1},
	\nonumber\\
	\Resp_a(0)
	&\equiv\lim_{\omega\rightarrow0}\Resp_a(\omega)
	=\frac{a}{4\pi^2}.
	\label{eq:response}
\end{align}
Within the weak-coupling Born--Markov and secular approximations, the reduced density matrix of the detector obeys~\cite{Tian:2024wby,Manzano:2020yyw}
\begin{equation}
	\begin{split}
		\dot{\rho}=\mathcal{L}[\rho]=&-i[H_{\mathrm D}+H_{\mathrm{LS}},\rho]+\sum_{i={\pm,z}}\gamma_i\Diss_{[\sigma_i]}\rho,
	\end{split}
	\label{eq:master}
\end{equation}
where $\Diss_{[L]}\rho=L\rho L^\dagger-\frac{1}{2}\{L^\dagger L,\rho\}$. The excitation and deexcitation rates are $\gamma_\pm=g^2\alpha^2\Resp_a(\pm\omega_0)$, respectively, while $\gamma_z=g^2\beta^2\Resp_a(0)$ is the zero-frequency dephasing coefficient. We neglect the Lamb shift term $H_{LS}$ because the corresponding correction to the detector energy gap is small compared with $\omega_0$. We use the inertial-limit spontaneous-emission rate, $\Gamma_0=g^2\alpha^2\omega_0/(2\pi)$, to define the dimensionless proper time $\Gamma_0\tau$. Since the dephasing channel does not affect the populations, the KMS relation $\Resp_a(+\omega_0)/	\Resp_a(-\omega_0)=e^{-2\pi\omega_0/a}$ implies a unique stationary state given by the Unruh Gibbs state~\cite{Arrechea:2021szl}
\begin{equation}
	\rhoU\equiv\rho_{\mathrm{ss}}
	=\frac{e^{-2\pi H_{\mathrm D}/a}}
	{\Tr e^{-2\pi H_{\mathrm D}/a}}.	\label{eq:gibbs}
\end{equation}

\begin{figure}[t]
	\includegraphics[width=0.8\linewidth]{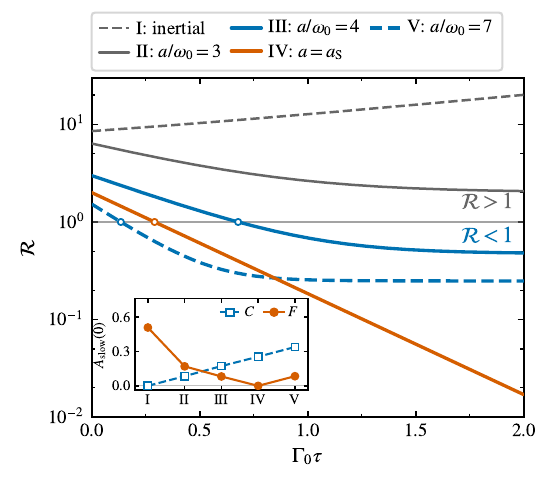}
	\caption{\textbf{Relaxation dynamics and suppression of the slow mode}. Cases I–V correspond to $a/\omega_0$=0,3,4,$a_{S}/\omega_0$,7, respectively, with $a_{S}/\omega_0=\pi/\operatorname{artanh}b$. Open circles mark crossings of $\mathcal{R}$=1. Inset: initial slow-subspace amplitudes $A_{{slow}}(0)$. m=0.9, b=0.54, and $\kappa=2$.}
	\label{fig:Ratio}
\end{figure}

\emph{Emergence of the strong Mpemba effect.---}
We write the detector state in Bloch form, $\rho=\frac{1}{2}\bigl(\mathbb{I}+\bm v\cdot\bm\sigma\bigr).$ The stationary state is \(\rho_{\mathrm U}=\frac12(\mathbb I-s\sigma_z)\), with \(s=\tanh(\pi \omega_0/a)\). The density matrix $\rho(\tau)$ admits the following mode decomposition
\begin{align}
	\rho(\tau)
	={\rho_{\mathrm U}}&+
	\frac{v_z(0)+s}{2}
	e^{-\gpop\tau}\sigma_z+\frac{v_-(0)}{2}
	e^{(-\gamma_{\mathrm{c}}
		-i\omega_0)\tau}\sigma_+
	\nonumber\\
	&+\frac{v_+(0)}{2}
	e^{(-\gamma_{\mathrm{c}}
		+i\omega_0)\tau}\sigma_- ,
\end{align}
where $v_\pm=v_x\pm i v_y$ and the population and coherence decay rates are respectively $\gpop=\gamma_+ +\gamma_-=\Gamma_0\coth(\pi \omega_0/a), \gcoh=\frac{\gpop}{2}+2\gamma_z.$ We quantify the distance between the state $\rho$ and the Unruh Gibbs state $\rho_U$ by the trace distance~\cite{Ares:2025onj,Nava:2025fdd}
\begin{equation}
	D_{T}(\rho(\tau))=\frac{1}{2}|\bm v (\tau) -\bm v_U|.
	\label{eq:distance}
\end{equation}
Two isospectral initial states can be constructed as 
\begin{equation}
	\bm v_C=(0,0,-m),\qquad
	\bm v_F=(\sqrt{m^2-b^2},0,-b),
	\label{eq:fixed_pair}
\end{equation}
where $0<b<m<1$. State $C$ is coherence-free, whereas $F$ has nonzero coherence. The labels denote the initially closer and farther states, respectively; their squared-distance difference at $\tau=0$ is $s(m-b)/2>0$ for finite acceleration.
\begin{figure}[t]
	\includegraphics[width=0.8\linewidth]{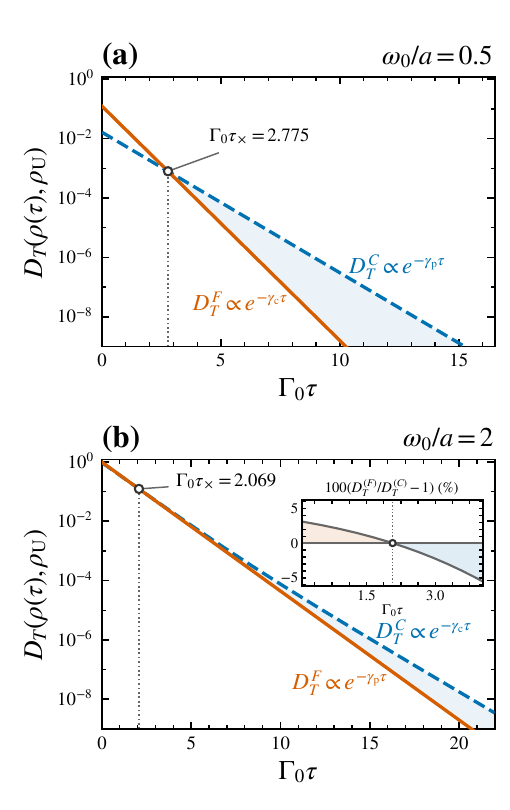}
	\caption{\textbf{Genuine strong-Mpemba crossings on a fixed unitary orbit.}
		Trace distance to $\rhoU$ for $\kappa=2$ and $m=0.95$.
		(a) At $\omega_0/a=0.50$. (b) At $\omega_0/a=2$ and $\theta=0.50$. The inset shows the signed relative difference $100({D_T^{F}}-{D_T^{C}})/{D_T^{C}}$ near the crossing. Dots and vertical lines denote the analytical crossings.}
	\label{fig:crossings}
\end{figure}
The difference 
\begin{equation}
	\begin{aligned}
		{D_T^{F}}^2(\tau)-{D_T^{C}}^2(\tau)=&\frac{m-b}{4}\big[(m+b)e^{-2\gcoh \tau}\\
		&+(2s-m-b)e^{-2\gpop \tau}\big]
	\end{aligned}
	\label{eq:fixed_difference}
\end{equation}
remains positive if $2s \geq m+b$. In the inertial limit, \(s\to1\) and \(\gamma_c=\gamma_p/2\), so the two trace distance curves for this fixed pair do not cross. Acceleration can enable a crossing by changing both the stationary state and the relative decay rates. Therefore, the two curves will cross at
\begin{equation}
	\tau_\times=\frac{\ln[(m+b)/(m+b-2s)]}{2(\gcoh-\gpop)},
	\label{eq:emergence_crossing}
\end{equation}
if and only if $\gcoh>\gpop$ and $2s<m+b$. The first condition makes population relaxation the slower channel, while the second ensures that \(F\) has a smaller population-mode amplitude than \(C\). For this fixed pair, the strong effect occurs at
\begin{equation}
	a_S=\frac{\pi \omega_0}{\atanh b},
\end{equation}
provided $\gcoh(a_S)>\gpop(a_S)$. Under this condition, the crossing onset precedes $a_S$. At $s=b$, the slow population component of $F$ vanishes, leaving relaxation at the faster rate $\gcoh$. Increasing the acceleration beyond $a_S$ restores the slow population component and yields an ordinary Mpemba crossing.

As shown in Fig~\ref{fig:ESME}, the black dash-dotted curve represents the onset of the Mpemba effect and the blue region indicates a positive time advantage. The color encodes \(\log_{10}[\tau_C(\varepsilon)/\tau_F(\varepsilon)]\), where \(\tau_i(\varepsilon)\) is the first time at which \(D_T^i(\tau)\leq\varepsilon\), thereby quantifying the time advantage afforded by the relaxation speedup.
Along the orange curve, exact cancellation of the slow-mode contribution yields a faster asymptotic decay. Away from this curve, a residual slow-mode contribution restores the slower asymptotic decay, although a finite-time relaxation advantage can persist.


The ratio of their trace distances from equilibrium is
\begin{equation}
	\mathcal R =\frac{D_{T}^F}{D_{T}^C}=\frac{\sqrt{(m^2-b^2)e^{-2(\gcoh-\gpop)\tau}+(s-b)^2}}{|m-s|}. \label{eq:ratio}
\end{equation}
We exclude \(m=s\), for which state $C$ coincides with \(\rho_U\) and the denominator in Eq.~(\ref{eq:ratio}) vanishes.
Fig.~\ref{fig:Ratio} illustrates these distinct behaviors for $b/m=0.6$. In inertial motion, $\mathcal R(\tau)$ increases and the initial distance ordering is preserved. When $\gamma_c>\gamma_p$, the ratio decreases toward $|s-b|/|m-s|$; a crossing occurs only when this limit is below unity. At $a=a_S\simeq5.2\omega_0$, the residual population amplitude vanishes and $\mathcal R(\tau)$ decays exponentially to zero. Thus, ordinary crossings retain the same asymptotic decay exponent for both states, whereas exact matching produces different exponents.

The asymptotic behavior in Fig.~\ref{fig:Ratio} is determined by the residual slow-mode amplitude. The inset connects these relaxation behaviors to the initial slow-mode amplitudes, which we define as the Hilbert--Schmidt norm of the contribution from the slowest-decaying sector
\begin{equation}
	A_{{slow}}(0)= 
	\begin{cases}
		\frac{1}{2}\sqrt{|v_+(0)|^2+|v_-(0)|^2},&\gamma_c<\gamma_p,\\[1mm] |v_z(0)+s|/\sqrt{2},&\gamma_p<\gamma_c. 
	\end{cases}
\end{equation}
Although the initial states remain fixed, acceleration changes both the stationary state and the slowest sector. In the population-limited regime, $A_{slow}^{c}(0)=|s-m|/\sqrt{2}$; it grows with acceleration only when $s<m$. At $a=a_S$, this amplitude vanishes for state $\bm v_F$, thereby causing a strong speedup in relaxation.

\emph{Acceleration-tuned spectral switch.--}
The relaxation bottleneck underlying the crossings above is determined by the Liouvillian spectrum, ${\it{spec}(\mathcal L)}=\left\{0,\,-\gpop,\,-\gcoh+i\omega_0,\,-\gcoh-i\omega_0\right\}$. The slowest nonstationary decay rate is therefore
$\Delta_{\mathcal L}=\min\{\gpop,\gcoh\}$, and the slowest relaxation sector is determined by
\begin{equation}
	\chi(a,\omega_0,\kappa)
	\equiv\frac{\gcoh}{\gpop}
	=\frac{1}{2}
	+\kappa\frac{\tanh(\pi \omega_0/a)}{\pi \omega_0/a},
	\qquad
	\kappa=\frac{\beta^2}{\alpha^2}.
	\label{eq:chi}
\end{equation}
The ratio approaches \(\chi=1/2\) in the inertial limit \(a\to0\), where coherence decays more slowly than population deviations. As acceleration increases, \(\chi\) approaches \(1/2+\kappa\), reflecting the growing contribution of longitudinal dephasing. For \(\chi>1\), the population mode \(\sigma_z\) is the slowest, whereas for \(\chi<1\), the complex-conjugate coherence pair \(\sigma_\pm\) forms the slowest sector. Since \(\chi\) increases monotonically with acceleration for \(\kappa>0\), a unique finite switch from coherence-limited to population-limited relaxation exists if and only if \(\kappa>1/2\). The unique positive solution satisfies
\begin{equation}
	\frac{\pi \omega_0}{a_*}\coth(\frac{\pi \omega_0}{a_*})=2\kappa.
	\label{eq:switch}
\end{equation}

\begin{figure}[t]
	\includegraphics[width=0.76\linewidth]{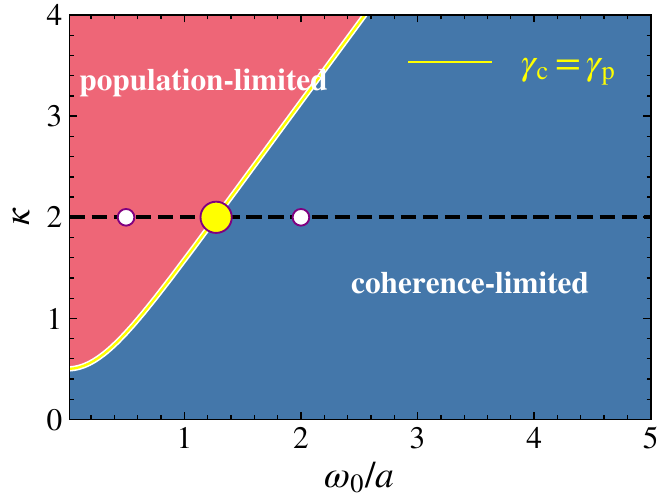}
	\caption{\textbf{Acceleration and coupling select the strong-Mpemba protocol.}
		Population-limited (red) and coherence-limited (blue) regimes in the $(\omega_0/a,\kappa)$ plane, separated by $\gamma_c=\gamma_p$ (yellow). The dashed line marks $\kappa=2$; the yellow dot marks the corresponding spectral switch. Open circles mark the parameters used in Fig.~\ref{fig:crossings}(a) (left) and (b) (right).}
	\label{fig:Phase}
\end{figure}

Fig.~\ref{fig:Phase} identifies the relaxation bottleneck across parameter space. The red and blue regions denote population-limited and coherence-limited relaxation, respectively. The yellow curve marks $\gcoh=\gpop\equiv\gamma$, where every trace distance decays as $D_T(\tau)=D_T(0)e^{-\gamma\tau}$ and no crossing occurs. In the inertial regime, relaxation is coherence limited. For fixed $\kappa>1/2$, increasing acceleration drives the detector across this boundary from the coherence-limited inertial regime into the population-limited regime.

As shown in Fig.~\ref{fig:sch}(b), the two parameter regimes admit distinct unitary transformations~\cite{Carollo2021,Moroder2024,z9t9-mwgv}, which eliminate the slowest modes and exponentially speed up the relaxation (see Supplemental Material for details)
\begin{equation}
	 \begin{cases}
		\bm v_C^{p}=(0,0,-m),  \bm v_F^{p}=(\sqrt{m^2-s^2},0,-s)\,&\chi>1,\\
		\bm v_C^{c}=(m\sin\theta,0,m\cos\theta), \bm v_F^{c}=(0,0,m) \,&\chi<1.
	\end{cases}\label{eq:cases}
\end{equation}
Here $0<m<1$ and $\theta\in(0,\pi)$; the population-limited protocol additionally requires $s<m$. In each regime, $F$ starts farther from equilibrium but lacks the slow component retained by $C$. It therefore overtakes $C$ and approaches equilibrium with the faster decay rate. The crossings marked in Fig.~\ref{fig:crossings} occur at
\begin{align}
		 &\tau_\times^{(p)}=\frac{1}{2(\gcoh-\gpop)}
		\ln\frac{m+s}{m-s}, \nonumber\\
		&\tau_\times^{(c)}=\frac{1}{2(\gpop-\gcoh)}
		\ln\frac{N}{m^2\sin^2\theta},\label{eq:crossing}
\end{align}
where \(N=(m+s)^2-(m\cos\theta+s)^2\).
 
For the fixed pair in Eq.~(\ref{eq:fixed_pair}), the inertial limit \(\chi=1/2\) yields no crossing. Once acceleration makes population relaxation the slower channel, the relevant slow-mode amplitude of $F$ is \(|s-b|/\sqrt{2}\). Its cancellation at \(s=b\) identifies this fixed state with the population-matched preparation in Eq.~(\ref{eq:cases}), producing the strong effect provided \(\chi(a_S)>1\). In the purely transverse limit, \(\kappa=0\), one instead has \(\gamma_c=\gamma_p/2\) at every acceleration~\cite{Wang:2025mpu}. The longitudinal channel is thus essential for the acceleration-induced switch underlying this fixed-pair mechanism.

\emph{Experimental prospects.---}
Direct experimental access to Unruh thermalization remains challenging, motivating proposals involving circular motion and quantum simulation~\cite{Zhou:2023xag,Zhou:2024vti,Peng:2024pfy,PhysRevLett.126.117401,Gooding:2020scc,Hegde:2018xub,Luo:2025iur,Hu:2018psq}. Mixed transverse--longitudinal coupling can arise in the energy eigenbasis of a coherently controlled detector~\cite{Liu:2025bzv,Liu:2025rks}, with an experimental counterpart in deep-strong-coupling circuit QED~\cite{Wang:2022mur}. Realizing our weak-coupling dynamics additionally requires the prescribed reservoir response.

Alternatively, the reduced dynamics can be simulated using a system ion, a resettable ancillary ion, and classical random control~\cite{Barreiro:2011fge}. Each time step $\delta\tau$ starts with independent preparation of an incoherent ancilla mixture with excitation probability $p_e=(1-s)/2$, where $s=\tanh(\pi\omega_0/a)$. The exchange gate
\begin{equation}
	U_{SA}=\exp[-i\vartheta(\sigma_+^S\sigma_-^A+\sigma_-^S\sigma_+^A)]
\end{equation}
can be synthesized from $\sigma_x\sigma_x$ and $\sigma_y\sigma_y$ M{\o}lmer--S{\o}rensen operations, with $0\leq\vartheta\leq\pi/2$ and $\cos^2\vartheta=e^{-\gamma_{\mathrm p}\delta\tau}$. Independent $\sigma_z$ operations with probability $q_z=(1-e^{-2\gamma_z\delta\tau})/2$ produce pure dephasing, and a $\sigma_z$ rotation through $\omega_0\delta\tau$ encodes free precession. These reduced channels commute and ideally implement $e^{\delta\tau\mathcal L_a}$ exactly at sampling times. Acceleration is encoded jointly in $p_e,\vartheta,q_z$, with the isospectral initial states, detector gap, and model couplings fixed. Independent calibration and Pauli tomography can probe crossings and, for $\gamma_c>\gamma_p$, suppression of the population-mode amplitude at $s=b$.

\emph{Conclusion.---}
Acceleration can induce ordinary and strong quantum Mpemba effects for a fixed pair of isospectral initial states undergoing Unruh thermalization. At a precisely matched acceleration, the initially farther state has zero overlap with the slowest decay mode and relaxes with a faster exponential rate. The mechanism combines an acceleration-dependent stationary state with competing population and coherence decay rates. For sufficiently strong longitudinal coupling, acceleration switches the slowest relaxation sector and thereby changes the required strong-Mpemba preparation. Our isospectral constructions identify the conditions for eliminating the slowest modes and achieving faster asymptotic relaxation in both the population- and coherence-limited regimes. Whether curvature-induced changes in the detector response permit similar spectral switching remains an open question.


\begin{acknowledgments}
	This work was supported by the National Natural Science Foundation of China under Grants No. 12475051, No. 12421005 and No. 12575050; the science and technology innovation Program of Hunan Province under grant No. 2024RC1050;  the innovative research group of Hunan Province under Grant No. 2024JJ1006; the Hangzhou Leading Youth Innovation and Entrepreneurship Team project  under Grant No. TD2024005; and the scientific research start-up funds of Hangzhou Normal University: 4245C50224204016.
\end{acknowledgments}

%

\clearpage
\onecolumngrid
\begin{center}
	{\large\bfseries Supplemental Material for ``Relativistic Strong Quantum Mpemba Effect''}
\end{center}
\setcounter{equation}{0}
\renewcommand{\theequation}{S\arabic{equation}}
\renewcommand{\theHequation}{supp.\arabic{equation}}
\setcounter{figure}{0}
\renewcommand{\thefigure}{S\arabic{figure}}
\renewcommand{\theHfigure}{supp.\arabic{figure}}
\setcounter{table}{0}
\renewcommand{\thetable}{S\arabic{table}}
\renewcommand{\theHtable}{supp.\arabic{table}}

\section{Liouville-space representation and modal decomposition}

In this section, we derive the Liouvillian spectrum and its
population and coherence eigenmodes directly from the master
equation. In the energy basis
$\{|e\rangle,|g\rangle\}$, defined by
\[
\sigma_z|e\rangle=|e\rangle,
\qquad
\sigma_z|g\rangle=-|g\rangle,
\]
the Hamiltonian is
\begin{equation}
	H_{\mathrm D}+H_{\mathrm{LS}}
	\approx\frac{\omega_0}{2}\sigma_z .
	\label{eq:sHeff}
\end{equation}
The reduced density matrix is parametrized as
\begin{equation}
	\rho=
	\begin{pmatrix}
		\rho_{ee} & \rho_{eg}\\
		\rho_{ge} & \rho_{gg}
	\end{pmatrix},
	\qquad
	\rho_{ee}+\rho_{gg}=1.
	\label{eq:sdensity}
\end{equation}

Substitution of Eq.~\eqref{eq:sdensity} into the master equation
gives
\begin{align}
	\dot{\rho}_{ee}
	&=-\gamma_- \rho_{ee}
	+\gamma_+\rho_{gg},
	\label{eq:spop-ee}\\
	\dot{\rho}_{gg}
	&=\gamma_- \rho_{ee}
	-\gamma_+\rho_{gg},
	\label{eq:spop-gg}\\
	\dot{\rho}_{eg}
	&=-\left(
	\frac{\gamma_++\gamma_- }{2}
	+2\gamma_z+i\omega_0
	\right)\rho_{eg},
	\label{eq:scoh-eg}\\
	\dot{\rho}_{ge}
	&=-\left(
	\frac{\gamma_++\gamma_- }{2}
	+2\gamma_z -i\omega_0
	\right)\rho_{ge}.
	\label{eq:scoh-ge}
\end{align}
The factor $2\gamma_z $ follows from
\begin{equation}
	\mathcal{D}[\sigma_z]\rho
	=\sigma_z\rho\sigma_z-\rho,
\end{equation}
whose off-diagonal matrix elements are
$-2\rho_{eg}$ and $-2\rho_{ge}$.

To obtain an explicit matrix representation of the Liouvillian, we
introduce the Liouville-space vector
\begin{equation}
	|\rho\rangle\!\rangle
	=
	\begin{pmatrix}
		\rho_{ee}&
		\rho_{eg}&
		\rho_{ge}&
		\rho_{gg}
	\end{pmatrix}^{\mathsf T}.
	\label{eq:svec}
\end{equation}
The master equation then takes the form
\begin{equation}
	\frac{d}{d\tau}|\rho\rangle\!\rangle
	=\mathcal{L}|\rho\rangle\!\rangle ,
	\label{eq:sliouville-equation}
\end{equation}
where
\begin{equation}
	\mathcal{L}=
	\begin{pmatrix}
		-\gamma_- 
		&0&0&\gamma_+\\
		0&-\gamma_{\mathrm{c}}-i\omega_0&0&0\\
		0&0&-\gamma_{\mathrm{c}}+i\omega_0&0\\
		\gamma_- 
		&0&0&-\gamma_+
	\end{pmatrix},
	\label{eq:sliouville-matrix}
\end{equation}
with
\begin{equation}
	\gpop
	\equiv\gamma_++\gamma_- ,
	\qquad
	\gamma_{\mathrm{c}}
	\equiv\frac{\gamma_++\gamma_- }{2}
	+2\gamma_z .
	\label{eq:srates}
\end{equation}

The population block of Eq.~\eqref{eq:sliouville-matrix} has
eigenvalues $0$ and $-\gpop$, whereas the two
coherence components are already diagonal. The complete Liouvillian
spectrum is therefore
\begin{equation}
	\operatorname{spec}(\mathcal L)
	=
	\left\{
	0,\,
	-\gpop,\,
	-\gamma_{\mathrm{c}}-i\omega_0,\,
	-\gamma_{\mathrm{c}}+i\omega_0
	\right\}.
	\label{eq:sspectrum}
\end{equation}
The corresponding right eigenoperators can be chosen as
\begin{align}
	R_0&=\rho_{\mathrm U},
	&\lambda_0&=0,\nonumber\\
	R_{\mathrm p}&=\sigma_z,
	&\lambda_{\mathrm p}&=-\gpop,\nonumber\\
	R_+&=\sigma_+,
	&\lambda_+&=-\gamma_{\mathrm{c}}
	-i\omega_0,\nonumber\\
	R_-&=\sigma_-,
	&\lambda_-&=-\gamma_{\mathrm{c}}
	+i\omega_0.
	\label{eq:seigenoperators}
\end{align}

The stationary populations follow from
$\dot{\rho}_{ee}=\dot{\rho}_{gg}=0$:
\begin{equation}
	\rho_{ee}^{\mathrm U}
	=\frac{\gamma_+}{\gpop},
	\qquad
	\rho_{gg}^{\mathrm U}
	=\frac{\gamma_- }{\gpop }.
	\label{eq:sstationary-populations}
\end{equation}
Using the KMS relation
\begin{equation}
	\frac{\gamma_+}{\gamma_- }
	=e^{-2\pi\omega_0/a},
\end{equation}
we obtain
\begin{equation}
	\rho_{ee}^{\mathrm U}
	=\frac{1}{1+e^{2\pi r}},
	\qquad
	\rho_{gg}^{\mathrm U}
	=\frac{e^{2\pi r}}{1+e^{2\pi r}},
	\qquad
	r=\frac{\omega_0}{a}.
	\label{eq:sstationary-explicit}
\end{equation}

The connection with the Bloch representation follows from
\begin{equation}
	\rho=\frac{1}{2}
	\left(\mathbb{I}+v_x\sigma_x+v_y\sigma_y+v_z\sigma_z\right),
	\label{eq:sbloch}
\end{equation}
for which
\begin{equation}
	\rho_{ee}=\frac{1+v_z}{2},
	\qquad
	\rho_{gg}=\frac{1-v_z}{2},
	\qquad
	\rho_{eg}=\frac{v_x-i v_y}{2}.
	\label{eq:sbloch-components}
\end{equation}
Using $e^{-x\sigma_z}=\cosh(x)\mathbb{I}
-\sinh(x)\sigma_z$, the stationary state can be written as
\begin{equation}
	\rho_{\mathrm U}
	=\frac{1}{2}\bigl[\mathbb{I}-s\sigma_z\bigr],
	\qquad
	s\equiv\tanh(\pi r).
\end{equation}
Its Bloch vector is therefore
$\bm v_{\mathrm U}=(0,0,-s)$.
Defining
\begin{equation}
	v_\pm\equiv v_x\pm i v_y,
\end{equation}
Eqs.~\eqref{eq:spop-ee}--\eqref{eq:scoh-ge} reduce to
\begin{equation}
	\dot v_z
	=-\gpop (v_z+s),
	\qquad
	\dot v_\pm
	=\left(
	-\gamma_{\mathrm{c}}
	\pm i\omega_0
	\right)v_\pm .
	\label{eq:sbloch-dynamics}
\end{equation}
Their exact solutions are
\begin{align}
	&v_z(\tau)+s
	=[v_z(0)+s]e^{-\gpop \tau},
	\label{eq:sbloch-pop-solution}\\
	&v_\pm(\tau)
	=v_\pm(0)
	e^{(-\gamma_{\mathrm{c}}
		\pm i\omega_0)\tau}.
	\label{eq:sbloch-coh-solution}
\end{align}

Equivalently, the deviation from the stationary state admits the exact Liouvillian-mode decomposition
\begin{align}
	\rho(\tau)-\rho_{\mathrm U}=\frac{v_z(0)+s}{2}	e^{-\gpop \tau}\sigma_z+\frac{v_x(0)-i v_y(0)}{2}	e^{(-\gamma_{\mathrm{c}}
		-i\omega_0)\tau}\sigma_+ +\frac{v_x(0)+i v_y(0)}{2} e^{(-\gamma_{\mathrm{c}}	+i\omega_0)\tau}\sigma_-. \label{eq:smodal-decomposition}
\end{align}
Equation~\eqref{eq:smodal-decomposition} makes explicit that the
initial state does not modify the Liouvillian eigenvalues. Instead,
initial-state control changes only the amplitudes multiplying the
corresponding eigenmodes.

Using the accelerated-vacuum response function, the two decay rates
become
\begin{align}
	\gpop=g^2\alpha^2
	\left[
	\Resp_a(+\omega_0)+\Resp_a(-\omega_0)
	\right]=g^2\alpha^2\frac{\omega_0}{2\pi}
	\coth(\pi r), \qquad \gamma_{\mathrm{c}} =\frac{\gpop}{2}
	+2g^2\beta^2\Resp_a(0) =\frac{\gpop}{2}
	+g^2\beta^2\frac{a}{2\pi^2}.
	\label{eq:sgamma-coh}
\end{align}
Consequently,
\begin{equation}
	\chi(r,\kappa)
	\equiv\frac{\gamma_{\mathrm{c}}}
	{\gpop}
	=\frac{1}{2}
	+\kappa\frac{\tanh(\pi r)}{\pi r},
	\qquad
	\kappa\equiv\frac{\beta^2}{\alpha^2}.
	\label{eq:schi}
\end{equation}

\section{Analytical derivation of the two strong-Mpemba protocols}
\label{sec:exact-crossings}

We provide here the complete analytical construction of the two strong-Mpemba protocols introduced in the main text. 

\subsection{Trace distance from the Unruh stationary state}

The Unruh stationary Bloch vector is
\begin{equation}
	\bm v_{\mathrm U}=(0,0,-s).
\end{equation}
For a general initial state $\rho(0)=\frac{1}{2}\left[\mathbb{I}+\bm v(0)\cdot\bm\sigma\right]$, the transverse rotation generated by $\omega_0$ leaves the transverse Bloch-vector magnitude
invariant, so that
\begin{equation}
	v_x^2(\tau)+v_y^2(\tau)
	=v_\perp^2(0)e^{-2\gcoh \tau},
	\qquad
	v_\perp^2(0)\equiv v_x^2(0)+v_y^2(0).
	\label{eq:s-transverse-radius}
\end{equation}

The deviation from the stationary state can be written as
\begin{equation}
	\rho(\tau)-\rhoU
	=\frac{1}{2}
	\left[\bm v(\tau)-\bm v_{\mathrm U}\right]
	\cdot\bm\sigma.
	\label{eq:s-state-deviation}
\end{equation}
Using the Pauli-matrix identity $\left(\bm d\cdot\bm\sigma\right)^2=|\bm d|^2\mathbb{I}$, the eigenvalues of $\rho(\tau)-\rhoU$ are $\mu_\pm(\tau)=\pm\frac{1}{2}\left|\bm v(\tau)-\bm v_{\mathrm U}\right|$. It follows that the trace distance is
\begin{align}
	D_{T}(\tau)\equiv	\frac{1}{2}
	\left\|\rho(\tau)-\rhoU\right\|_1=\frac{1}{2}
	\sqrt{
		v_\perp^2(0)e^{-2\gcoh \tau}
		+[v_z(0)+s]^2e^{-2\gpop \tau}
	}.
	\label{eq:s-trace-general}
\end{align}
Equation~\eqref{eq:s-trace-general} separates explicitly the
coherence and population contributions to the distance from
equilibrium.

\subsection{Population-limited protocol}

Suppose that
\begin{equation}
	\chi=\frac{\gcoh}{\gpop}>1,
\end{equation}
so that $\gpop<\gcoh$ and the population mode is the unique slowest decay mode. Choose a Bloch radius $m$ satisfying
\begin{equation}
	0<s<m\leq1,
	\label{eq:s-pop-domain}
\end{equation}
and consider the two initial states
\begin{equation}
	\bm v_C=(0,0,-m),
	\qquad
	\bm v_F=
	\left(\sqrt{m^2-s^2},0,-s\right).
	\label{eq:s-pop-states}
\end{equation}
Both states are physical because their Bloch-vector lengths satisfy $|\bm v_C|=|\bm v_F|=m\leq1$. Their density matrices consequently have the same eigenvalues,
\begin{equation}
	\lambda_\pm=\frac{1\pm m}{2},
	\label{eq:s-isospectral-eigenvalues}
\end{equation}
and are related by a unitary rotation. For example, defining
\begin{equation}
	\varphi_p=\arccos\left(\frac{s}{m}\right),
	\qquad
	U_p=e^{+i\varphi_p\sigma_y/2},
\end{equation}
one has
\begin{equation}
	\rho_F=U_p\rho_CU_p^\dagger.
\end{equation}

For the reference state $C$,
\begin{equation}
	v_{\perp,C}(0)=0,
	\qquad
	v_{z,C}(0)+s=-(m-s),
\end{equation}
and Eq.~\eqref{eq:s-trace-general} gives
\begin{equation}
	D_C^{(p)}(\tau)
	=\frac{m-s}{2}e^{-\gpop \tau}.
	\label{eq:s-pop-distance-C}
\end{equation}

Consequently, its distance from equilibrium decays at the slow population rate \(\gamma_{\rm p}\).

By contrast, for the optimized state $F$,
\begin{equation}
	v_{\perp,F}(0)=\sqrt{m^2-s^2},
	\qquad
	v_{z,F}(0)+s=0.
\end{equation}
The slow population mode is therefore absent from its relaxation
trajectory, and the remaining coherent deviation decays at the
faster rate, 
\begin{equation}
	D_F^{(p)}(\tau)
	=\frac{\sqrt{m^2-s^2}}{2}e^{-\gcoh \tau}.
	\label{eq:s-pop-distance-F}
\end{equation}
Although $F$ contains no slow population component, it is initially
farther from equilibrium:
\begin{align}
	\frac{D_F^{(p)}(0)}{D_C^{(p)}(0)}=	\frac{\sqrt{m^2-s^2}}{m-s}=\sqrt{\frac{m+s}{m-s}}>1.
	\label{eq:s-pop-initial-order}
\end{align}

At arbitrary time,
\begin{equation}
	\frac{D_F^{(p)}(\tau)}{D_C^{(p)}(\tau)}
	=
	\sqrt{\frac{m+s}{m-s}}\,
	e^{-(\gcoh-\gpop)\tau}.
	\label{eq:s-pop-distance-ratio}
\end{equation}
Because $\gcoh-\gpop>0$, this ratio decreases strictly from a value
larger than one to zero. Consequently, the two trace-distance curves have exactly one finite-time crossing. Solving
$D_F^{(p)}(\tau_\times)=D_C^{(p)}(\tau_\times)$ yields
\begin{equation}
	\tau_\times^{(p)}
	=
	\frac{1}{2(\gcoh-\gpop)}
	\ln\left(\frac{m+s}{m-s}\right).
	\label{eq:s-crossing-pop}
\end{equation}
Before $\tau_\times^{(p)}$, the optimized state is farther from
equilibrium, whereas after $\tau_\times^{(p)}$ it remains closer and
decays with the faster coherence rate.

\subsection{Coherence-limited protocol}

We next consider
\begin{equation}
	\chi=\frac{\gcoh}{\gpop}<1,
\end{equation}
for which $\gcoh<\gpop $ and the complex-conjugate coherence pair constitutes the slowest Liouvillian sector. For finite acceleration, $0<s\leq1$. Let
\begin{equation}
	0<m\leq1,
	\qquad
	0<\theta<\pi,
	\label{eq:s-coh-domain}
\end{equation}
and choose
\begin{equation}
	\bm v_C=
	\left(m\sin\theta,0,m\cos\theta\right),
	\qquad
	\bm v_F=(0,0,m).
	\label{eq:s-coh-states}
\end{equation}
Again,
\begin{equation}
	|\bm v_C|=|\bm v_F|=m,
\end{equation}
so the two states have the common eigenvalues in
Eq.~\eqref{eq:s-isospectral-eigenvalues}. They are connected, for
example, by
\begin{equation}
	U_c=e^{+i\theta\sigma_y/2},
	\qquad
	\rho_F=U_c\rho_CU_c^\dagger.
\end{equation}

The optimized state $F$ is diagonal in the detector energy basis:
\begin{equation}
	v_{F,x}(0)=v_{F,y}(0)=0.
\end{equation}
It therefore has zero projection onto both slow coherence modes.
Its trace distance is
\begin{equation}
	D_F^{(c)}(\tau)
	=\frac{m+s}{2}e^{-\gpop \tau}.
	\label{eq:s-coh-distance-F}
\end{equation}
In contrast, the reference state $C$ has a nonzero coherence component and, in general, a population deviation, giving
\begin{align}
	\left[D_C^{(c)}(\tau)\right]^2	=\frac{1}{4}\biggl[m^2\sin^2\theta\,e^{-2\gcoh \tau}+(m\cos\theta+s)^2e^{-2\gpop \tau}\biggr].
	\label{eq:s-coh-distance-C}
\end{align}

At $\tau=0$, their squared-distance difference is
\begin{align}
	\left[D_F^{(c)}(0)\right]^2
	-\left[D_C^{(c)}(0)\right]^2
	&=
	\frac{1}{4}
	\bigl[
	(m+s)^2
	-m^2\sin^2\theta 
	-(m\cos\theta+s)^2
	\bigr]=\frac{ms}{2}(1-\cos\theta)>0.
	\label{eq:s-coh-initial-order}
\end{align}
Thus the coherence-free optimized state is initially farther from equilibrium. Define $	N\equiv	(m+s)^2-(m\cos\theta+s)^2$. The time-dependent squared-distance difference becomes
\begin{align}
	\left[D_F^{(c)}(\tau)\right]^2
	-\left[D_C^{(c)}(\tau)\right]^2
	=\frac{1}{4}\biggl[N e^{-2\gpop \tau}-m^2\sin^2\theta\,e^{-2\gcoh \tau}
	\biggr].
	\label{eq:s-coh-distance-difference}
\end{align}
The crossing condition is therefore
\begin{equation}
	N e^{-2\gpop \tau_\times}
	=
	m^2\sin^2\theta\,
	e^{-2\gcoh \tau_\times}.
\end{equation}
Solving for $\tau_\times$ gives
\begin{equation}
	\tau_\times^{(c)}
	=
	\frac{1}{2(\gpop-\gcoh)}
	\ln\left(
	\frac{N}{m^2\sin^2\theta}
	\right)
	\label{eq:s-crossing-coh}
\end{equation}

The logarithm is strictly positive. Indeed,
\begin{equation}
	N-m^2\sin^2\theta
	=2ms(1-\cos\theta)>0,
	\label{eq:s-N-positivity}
\end{equation}
and hence $	\frac{N}{m^2\sin^2\theta}>1.$ For $0<\theta<\pi$, the same ratio can also be written as
\begin{equation}
	\frac{N}{m^2\sin^2\theta}=1+\frac{2s}{m(1+\cos\theta)}.
	\label{eq:s-N-simplified}
\end{equation}
Because $\gpop>\gcoh$, Eq.~\eqref{eq:s-coh-distance-difference}
changes sign exactly once. The crossing in
Eq.~\eqref{eq:s-crossing-coh} is therefore finite, positive, and
unique.

\section{Role of coherence in relaxation processes}

The distinction between population- and coherence-limited relaxation also clarifies the role of coherence in initial-state preparation~\cite{Ivander:2023asg,Longhi:2024pko,Summer:2025wsa}. Coherence in the detector energy basis is quantified by the $l_1$ norm~\cite{Baumgratz:2013ecx}
\begin{equation}
	\mathcal{C}=\sum_{i\neq j}|\rho_{ij}|=\sqrt{v_x^2+v_y^2}.
\end{equation}
In terms of the initial coherence, the trace distance is given by
\begin{equation}
	D_T(\tau)=\frac{1}{2}\sqrt{\mathcal C(0)^2 e^{-2\gcoh \tau}+(v_z(0)+s)^2 e^{-2\gpop \tau}}.
\end{equation}

For the state $\bm v_F$ constructed in Eq.~(\ref{eq:fixed_pair}), in the coherence-limited region, the slowest-mode amplitude $A_{slow}^c(0)=\mathcal C(0)/\sqrt{2}$, which increases linearly with coherence.
In contrast, in the population-limited region it becomes $A_{{slow}}^p(\mathcal C) =\left|s-\sqrt{m^2-\mathcal C^2}\right|/\sqrt{2}$. 

\begin{figure*}[t]
	\includegraphics[width=0.8\linewidth]{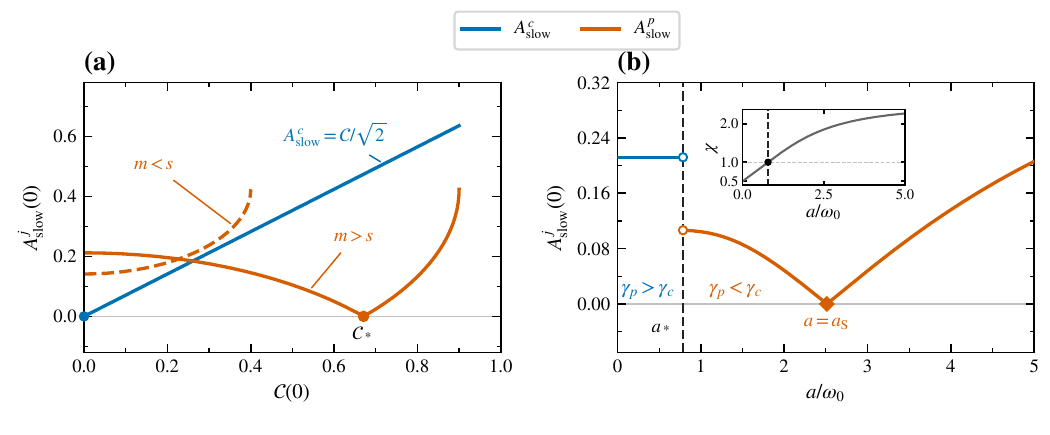}
	\caption{\textbf{Coherence and acceleration control the slow-mode amplitude \(A_{{slow}}^{j}(0)\) (\(j=c,p\)).} (a) Coherence dependence at \(s=0.6\), with population branches for \(m=0.9\) (solid) and \(m=0.4\) (dashed). (b) Acceleration-selected slow-sector amplitude for \(m=0.9\), \(\mathcal C(0)=0.3\), and \(\kappa=2\). Open circles denote one-sided limits at \(\gamma_c=\gamma_p\); the diamond marks population slow-mode cancellation. Inset: the rate ratio \(\chi=\gamma_c/\gamma_p\) crosses unity at the spectral switch \(a_*\).}
	\label{fig:Coh}
\end{figure*}
Fig.~\ref{fig:Coh} illustrates coherence control of the mode amplitudes and acceleration-dependent selection of the slow sector. In panel (a), the population amplitude \(A_{{slow}}^p\) decreases to zero at \(\mathcal C_*=\sqrt{m^2-s^2}\) and subsequently increases for \(m>s\), whereas it increases monotonically for \(m<s\). The coherence amplitude \(A_{{slow}}^c\) increases linearly. In panel (b), increasing acceleration at fixed \(\omega_0\) lowers \(s=\tanh(\pi\omega_0/a)\) and switches the slow sector from coherence to population at \(\gamma_c=\gamma_p\). For the parameters in panel (b), the slow-mode amplitude vanishes within the population-limited regime at $s_*=\sqrt{m^2-\mathcal C(0)^2}$, corresponding to $a_S=\pi\omega_0/\atanh(s_*)$. The unequal one-sided limits at $a_*$ reflect the change of the selected slow sector; the state and its trace distance remain continuous. For nonzero initial coherence, this cancellation leaves relaxation governed by the faster coherence modes, providing the spectral mechanism for the strong Mpemba effect.

\section{Experimental implementation of the encoded detector dynamics}
\label{sec:expapp}

We describe a digital protocol for implementing the reduced dynamics of
the accelerated detector using a system ion, a resettable ancillary ion,
and classical randomization. The construction uses the coherent gates
and optical-pumping reset demonstrated in the trapped-ion toolbox of
Barreiro et al.~\cite{Barreiro:2011fge}. The ancillary ion provides a
reusable environmental degree of freedom, while its reinitialization
removes system--ancilla correlations between successive cycles.
Acceleration is represented by a prescribed family of quantum channels
whose parameters reproduce the detector relaxation rates.

\subsection{Target dynamics and control parameters}
\label{subsec:expapp-target}

Following the main text, we neglect the Lamb shift and consider
\begin{equation}
	\begin{aligned}
		\dot{\rho}
		={}-i\left[\frac{\omega_0}{2}\sigma_z,\rho\right]	+\gamma_+\mathcal{D}[\sigma_+]\rho+\gamma_-\mathcal{D}[\sigma_-]\rho
		+\gamma_z\mathcal{D}[\sigma_z]\rho
		\equiv\mathcal{L}[\rho],
	\end{aligned}
	\label{eq:expapp-master}
\end{equation}
where$	\mathcal{D}[L]\rho	=L\rho L^\dagger-\frac12\{L^\dagger L,\rho\}.$ The system-ion states $|e\rangle_S$ and $|g\rangle_S$ encode the detector energy eigenstates, with $\sigma_z=|e\rangle\langle e|-|g\rangle\langle g|$. For the scalar-field response considered here, the rates are
\begin{equation}
	\begin{aligned}
		\gamma_{\mathrm p}
		&\equiv\gamma_++\gamma_-
		=\Gamma_0\coth\left(\frac{\pi\omega_0}{a}\right),\\
		\gamma_+
		&=\frac{\gamma_{\mathrm p}}{2}(1-s),\\
		\gamma_-
		&=\frac{\gamma_{\mathrm p}}{2}(1+s),\\
		\gamma_z
		&=\Gamma_0\kappa\frac{a}{2\pi\omega_0},
	\end{aligned}
	\label{eq:expapp-rates}
\end{equation}
where
\begin{equation}
	s=\tanh\left(\frac{\pi\omega_0}{a}\right),\qquad
	\Gamma_0=\frac{g^2\alpha^2\omega_0}{2\pi},\qquad
	\kappa=\frac{\beta^2}{\alpha^2}.
\end{equation}
In the detector energy basis, the target evolution over a simulated
proper-time interval $\delta\tau$ is
\begin{equation}
	\begin{aligned}
		\rho_{ee}(\tau+\delta\tau)
		&=p_e+(\rho_{ee}(\tau)-p_e)e^{-\gamma_{\mathrm p}\delta\tau},\\
		\rho_{gg}(\tau+\delta\tau)
		&=1-\rho_{ee}(\tau+\delta\tau),\\
		\rho_{eg}(\tau+\delta\tau)
		&=e^{-\gamma_{\mathrm c}\delta\tau}
		e^{-i\omega_0\delta\tau}\rho_{eg}(\tau),\\
		\rho_{ge}(\tau+\delta\tau)
		&=\rho_{eg}(\tau+\delta\tau)^*,
	\end{aligned}
	\label{eq:expapp-target-map}
\end{equation}
with
\begin{equation}
	p_e=\frac{\gamma_+}{\gamma_{\mathrm p}}
	=\frac{1-s}{2},
	\qquad
	\gamma_{\mathrm c}
	=\frac{\gamma_{\mathrm p}}{2}+2\gamma_z.
\end{equation}
Equation~\eqref{eq:expapp-target-map} determines the required ancilla
preparation, exchange angle, phase-flip probability, and coherent
rotation.

\subsection{Joint implementation of excitation and deexcitation}
\label{subsec:expapp-exchange}

At the beginning of every cycle, the ancilla is independently prepared
in the incoherent mixture
\begin{equation}
	\rho_A
	=(1-p_e)|g\rangle_A\langle g|
	+p_e|e\rangle_A\langle e|.
	\label{eq:expapp-ancilla}
\end{equation}
This state can be obtained by resetting the ancilla to $|g\rangle_A$
and applying a bit flip with probability $p_e$. The random choice is
renewed in every cycle, and all outcomes are included in the
unconditional ensemble. A coherent superposition with the same
populations would not, by itself, implement the diagonal ancilla state
in Eq.~\eqref{eq:expapp-ancilla}.

The system and ancilla then undergo the partial excitation exchange
\begin{equation}
	U_{SA}(\vartheta)
	=\exp\left[
	-i\vartheta
	\left(
	\sigma_+^S\sigma_-^A+
	\sigma_-^S\sigma_+^A
	\right)
	\right],
	\label{eq:expapp-exchange-unitary}
\end{equation}
where $0\leq\vartheta<\pi/2$. Its action on the single-excitation
subspace is
\begin{equation}
	\begin{aligned}
		|e,g\rangle
		&\longmapsto
		\cos\vartheta\,|e,g\rangle
		-i\sin\vartheta\,|g,e\rangle,\\
		|g,e\rangle
		&\longmapsto
		\cos\vartheta\,|g,e\rangle
		-i\sin\vartheta\,|e,g\rangle,
	\end{aligned}
\end{equation}
while $|e,e\rangle$ and $|g,g\rangle$ remain unchanged.
Tracing out the ancilla gives
\begin{equation}
	\begin{aligned}
		\mathcal{A}_{p_e,\vartheta}(\rho)
		=\operatorname{Tr}_A\big[
		U_{SA}(\vartheta)
		(\rho\otimes\rho_A)
		U_{SA}^\dagger(\vartheta)
		\big].
	\end{aligned}
	\label{eq:expapp-gad}
\end{equation}
The reduced matrix elements satisfy
\begin{equation}
	\begin{aligned}
		\rho_{ee}(\tau+\delta\tau)
		&=p_e+(\rho_{ee}(\tau)-p_e)\cos^2\vartheta,\\
		\rho_{eg}(\tau+\delta\tau)
		&=\rho_{eg}(\tau)\cos\vartheta.
	\end{aligned}
	\label{eq:expapp-gad-elements}
\end{equation}

The ancilla population determines the relative excitation and
deexcitation weights, whereas $\sin^2\vartheta$ determines the total
exchange probability. Matching Eq.~\eqref{eq:expapp-gad-elements}
to the target population evolution gives
\begin{equation}
		p_e=\frac{\gamma_+}{\gamma_{\mathrm p}},
		\qquad
		\cos^2\vartheta
		=e^{-\gamma_{\mathrm p}\delta\tau}.
	\label{eq:expapp-exchange-controls}
\end{equation}
Equivalently,
\begin{equation}
	\vartheta
	=\arccos\left(
	e^{-\gamma_{\mathrm p}\delta\tau/2}
	\right).
\end{equation}
The same gate automatically produces the coherence factor $\cos\vartheta=e^{-\gamma_{\mathrm p}\delta\tau/2}$, reproducing the contribution $\gamma_{\mathrm p}/2$ to the coherence-decay rate.

The finite-step transition probabilities are
\begin{equation}
	\begin{aligned}
		P_{g\rightarrow e}
		&=p_e\left(1-e^{-\gamma_{\mathrm p}\delta\tau}\right),\\
		P_{e\rightarrow g}
		&=(1-p_e)\left(1-e^{-\gamma_{\mathrm p}\delta\tau}\right).
	\end{aligned}
\end{equation}
In the short-step limit, these become $\gamma_+\delta\tau$ and
$\gamma_-\delta\tau$, respectively. The exponential prescription in
Eq.~\eqref{eq:expapp-exchange-controls} also matches the exact
finite-time evolution.

The exchange gate can be synthesized from $\sigma_x\sigma_x$ and $\sigma_y\sigma_y$ operations:
\begin{equation}
	U_{SA}(\vartheta)
	=e^{-i\vartheta \sigma_x^S \sigma_x^A/2}
	e^{-i\vartheta \sigma_y^S \sigma_y^A/2}.
	\label{eq:expapp-exchange-decomposition}
\end{equation}
This factorization is exact because $\sigma_x^S \sigma_x^A$ commutes with $\sigma_y^S \sigma_y^A$. For two ions, the collective M{\o}lmer--S{\o}rensen gate
\begin{equation}
	U_{X^2}(\zeta)
	=\exp\left[
	-\frac{i\zeta}{4}(\sigma_x^S+\sigma_x^A)^2
	\right]
\end{equation}
is equivalent, up to a global phase, to $e^{-i\zeta X_SX_A/2}$. The corresponding $\sigma_y$ operation is obtained by changing the drive phase. Two appropriately calibrated collective operations therefore implement Eq.~\eqref{eq:expapp-exchange-decomposition}.
%

\subsection{Pure dephasing and coherent precession}
\label{subsec:expapp-dephasing}

The longitudinal dissipator is implemented by applying $\sigma_z$
with probability $q_z$:
\begin{equation}
	\mathcal{P}_{q_z}(\rho)
	=(1-q_z)\rho+q_z\sigma_z\rho\sigma_z.
\end{equation}
This channel preserves the populations and transforms the coherence as
\begin{equation}
	\rho_{eg}(\tau+\delta\tau)=(1-2q_z)\rho_{eg}(\tau).
\end{equation}
Since
\begin{equation}
	\left[\mathcal{D}[\sigma_z]\rho\right]_{eg}
	=-2\rho_{eg},
\end{equation}
the required probability is
\begin{equation}
		q_z=\frac{1-e^{-2\gamma_z\delta\tau}}{2}.
	\label{eq:expapp-phase-flip}
\end{equation}
The phase-flip choice is independent between cycles and independent of the random choice used for ancilla preparation. A $Z$ rotation
through $\pi$ implements $\sigma_z$ up to a global phase.

The detector Hamiltonian is represented by
\begin{equation}
	U_D=e^{-i\varphi_D\sigma_z/2},
	\qquad
	\varphi_D=\omega_0\delta\tau.
	\label{eq:expapp-precession}
\end{equation}
Known native phases are compensated or tracked in a suitable rotating frame. The theoretical gap $\omega_0$ is therefore encoded in the
logical phase accumulated per simulated step. 

The complete reduced channel is
\begin{equation}
	\mathcal{E}_a(\delta\tau)
	=\mathcal{U}_D
	\circ\mathcal{P}_{q_z}
	\circ\mathcal{A}_{p_e,\vartheta},
\end{equation}
where $\mathcal{U}_D(\rho)=U_D\rho U_D^\dagger$.
Its coherence factor is
\begin{equation}
	\begin{aligned}
		&\cos\vartheta(1-2q_z)e^{-i\varphi_D}\\
		&\qquad=
		e^{-(\gamma_{\mathrm p}/2+2\gamma_z)\delta\tau}
		e^{-i\omega_0\delta\tau}.
	\end{aligned}
\end{equation}
Together with the population map, this establishes
\begin{equation}
		\mathcal{E}_a(\delta\tau)
		=e^{\delta\tau\mathcal{L}_a}.
	\label{eq:expapp-exact-map}
\end{equation}
The complete generalized amplitude-damping channel, phase-flip channel, and $\sigma_z$ rotation commute. Their ideal composition therefore reproduces the target dynamics exactly at the sampling times, without a time-splitting approximation between these three reduced channels.

\subsection{Reset, timing, and acceleration encoding}
\label{subsec:expapp-timing}

Following each exchange, the ancilla is reset and independently
reprepared before the next interaction. Ideally, this implements
\begin{equation}
	\rho_{SA}
	\longmapsto
	\operatorname{Tr}_A(\rho_{SA})\otimes\rho_A,
\end{equation}
so that correlations from one interaction do not enter the next cycle. Optical pumping of the ancilla must preserve the system state, using individual control or suitable protection operations.

After $n$ cycles,
\begin{equation}
	\tau_n=n\delta\tau,
	\qquad
	\rho_n=e^{\tau_n\mathcal{L}_a}\rho_0.
\end{equation}
If each physical cycle takes $T_{\mathrm{cyc}}$, the laboratory time is $t_{\mathrm{lab}}=nT_{\mathrm{cyc}}$. This duration includes preparation, gate operations, and reset; it need not equal $\delta\tau$.

Introducing
\begin{equation}
	\mathcal A=\frac{a}{\omega_0},
	\qquad
	h=\Gamma_0\delta\tau,
\end{equation}
the controls become
\begin{equation}
		\begin{aligned}
			&p_e(\mathcal A)
			=\frac{1-\tanh(\pi/\mathcal A)}{2},\\
			&\vartheta(\mathcal A,h)
			=\arccos\left[
			e^{-h\coth(\pi/\mathcal A)/2}
			\right],\\
			&q_z(\mathcal A,h,\kappa)
			=\frac{1-e^{-\kappa \mathcal A h/\pi}}{2},\\
			&\varphi_D
			=\frac{\omega_0}{\Gamma_0}h.
		\end{aligned}
	\label{eq:expapp-dimensionless-controls}
\end{equation}
The inertial limit $\mathcal A\rightarrow0$ gives $p_e=0$, $\cos^2\vartheta=e^{-h}$, and $q_z=0$.

The microscopic combinations $g^2\alpha^2$ and $g^2\beta^2$ enter through the transition and dephasing strengths, respectively. The former determines the exchange setting through $\Gamma_0$, while the latter determines the phase-flip probability through $\gamma_z$. The ratio $\kappa$ specifies their relative strength. Changing acceleration requires a coordinated update of $p_e,\vartheta,q_z$; changing only the ancilla population would generally not reproduce the prescribed detector dynamics.

The stationary populations satisfy
\begin{equation}
	\frac{p_e}{1-p_e}
	=\frac{\gamma_+}{\gamma_-}
	=e^{-2\pi\omega_0/a},
\end{equation}
and hence
\begin{equation}
	\begin{aligned}
		\rho_{\mathrm{ss}}
		&=p_e|e\rangle\langle e|
		+(1-p_e)|g\rangle\langle g|\\
		&=\frac12(I-s\sigma_z)
		=\frac{e^{-2\pi H_D/a}}
		{\operatorname{Tr}e^{-2\pi H_D/a}}.
	\end{aligned}
\end{equation}
The additional dephasing channel changes the coherence-decay rate without modifying these stationary populations.

\subsection{Initial-state preparation and acceleration scan}
\label{subsec:expapp-preparation}

For the fixed isospectral pair,
\begin{equation}
	\begin{aligned}
		\boldsymbol v_C=(0,0,-m),\qquad
		\boldsymbol v_F=(\sqrt{m^2-b^2},0,-b),
	\end{aligned}
	\qquad
	0<b<m<1,
	\label{eq:expapp-initial-pair}
\end{equation}
the common mixed-state seed
\begin{equation}
	\rho_C=\frac12(I-m\sigma_z)
\end{equation}
is prepared with system excitation probability $(1-m)/2$. Applying
\begin{equation}
	U_F=e^{+i\phi\sigma_y/2},
	\qquad
	\phi=\arccos(b/m),
\end{equation}
produces $\rho_F=U_F\rho_CU_F^\dagger$, preserving the eigenvalues and purity.

The parameters $m,b,\omega_0,\kappa,\Gamma_0$, and $\delta\tau$ remain fixed throughout the acceleration scan. Each trajectory uses a constant encoded acceleration. The strong-matching point is
\begin{equation}
	\mathcal A_S=\frac{\pi}{\operatorname{artanh}b},
\end{equation}
where $s=b$ and the initial population-mode amplitude of $F$ vanishes. To make population relaxation the slower sector at this point, one selects
\begin{equation}
	\gamma_{\mathrm c}(\mathcal A_S)>
	\gamma_{\mathrm p}(\mathcal A_S),
\end{equation}
or, equivalently,
\begin{equation}
	\kappa>
	\frac{\operatorname{artanh}b}{2b}.
\end{equation}
The scan can then cover a regime without crossing, an ordinary crossing, the matched strong effect, and the recovery of a residual slow tail away from matching.

\end{document}